\documentclass{llncs} 

\usepackage{amsmath}
\usepackage{bbm}
\usepackage{mathtools}
\usepackage{amssymb}
\usepackage{graphicx}
\usepackage{color}
\usepackage{latexsym}
\usepackage[utf8]{inputenc}
\usepackage[T1]{fontenc}
\usepackage{comment}
\usepackage[english]{babel}
\usepackage{stmaryrd}
\usepackage{amssymb} 
\usepackage[mathcal]{eucal}
\usepackage{fancybox}
\usepackage{algorithm,algorithmicx,algpseudocode}
\usepackage{xcolor}
\usepackage{palatino}
\usepackage[colorinlistoftodos,textsize=small]{todonotes}
\usepackage{framed} 
\usepackage{cases} 
\usepackage{relsize}

\newtheorem{prop}{Proposition}

\newtheorem{rem}{Remark}
\newtheorem{cor}[prop]{Corollary}

\newcommand{\ShowTODO}[1]{{#1}}
\renewcommand{\ShowTODO}[1]{}

\newcommand{\TODOB}[2]{\ShowTODO{\todo[linecolor=#1,inline, backgroundcolor=#1!60!white,bordercolor=#1]{\sf #2}}}

\newcommand{\TODOAll}[1]{\TODOB{pink}{{\bfseries TODO All:} #1}}

\newcommand{\TODOCedric}[1]{\TODOB{green!70!black}{{\bfseries TODO Cédric:} #1}}

\definecolor{darkgray}{gray}{0.50}

\def\emm#1,{{\em #1}}
\newcommand{\beq}{\begin{equation}}
\newcommand{\eeq}{\end{equation}}

\DeclareMathOperator{\Ins}{Ins}
\DeclareMathOperator{\one}{\boldsymbol 1}
\DeclareMathOperator{\several}{\boldsymbol +}
\DeclareMathOperator{\Del}{Del}

\DeclareMathOperator{\PrA}{Pr}

\DeclareMathOperator{\DelRoot}{DelRoot}
\DeclareMathOperator{\InsRoot}{InsRoot}
\DeclareMathOperator{\MatchRoot}{MatchRoot}

\DeclareMathOperator*{\Union}{\bigoplus}

\newcommand{\Insr}[1]{\InsRoot\left(#1\right)}
\newcommand{\Delr}[1]{\DelRoot\left(#1\right)}
\newcommand{\Matchr}[1]{\MatchRoot\left(#1\right)}
\newcommand{\Insrr}[2]{\InsRoot\left(#1,#2\right)}
\newcommand{\Delrr}[2]{\DelRoot\left(#1,#2\right)}
\newcommand{\Matchrr}[2]{\MatchRoot\left(#1,#2\right)}

\newcommand{\Vset}[1]{\mathcal V^{#1}}
\newcommand{\VHset}{\mathcal {VH}}
\newcommand{\Hset}[3]{\mathcal {H}_{#1,#2,#3}}
\newcommand{\Hcut}[4]{\mathcal {\overline H}^{\ #3,#4}_{#1,#2}}

\algrenewcommand\algorithmicindent{2.5em}
\algrenewcommand{\algorithmiccomment}[1]{{\color{brown} \hfill$\blacktriangleright$ #1 }}

\newcommand{\InsDelMode}{{\sf I|D}}
\newcommand{\DelMode}{{\sf D}}
\newcommand{\ccsimeq}{=}

\newcommand{\VMMO}[2]{\text{\bf [VIREZ MOI]}}
\newcommand{\VMOM}[2]{\text{\bf [VIREZ MOI]}}

\title{Counting, generating and sampling tree alignments}

\author{Cedric Chauve\inst{1}
\and Julien Courtiel\inst{1,2}
\and Yann Ponty\inst{2,3}}

\institute{%
Department of Mathematics, Simon Fraser University
\and Pacific Institute for the Mathematical Sciences
\and CNRS-LIX, Ecole Polytechnique}

\begin{document}
\maketitle

\vspace*{-5mm}
\begin{abstract}
  Pairwise ordered tree alignment are combinatorial objects that appear in RNA secondary structure comparison. However, the usual representation of tree alignments as supertrees is ambiguous, i.e. two distinct supertrees may induce identical sets of matches between identical pairs of trees. This ambiguity is uninformative, and detrimental to any probabilistic analysis.

  In this work, we consider tree alignments up to equivalence. Our first result is a precise asymptotic  enumeration of tree alignments, obtained from a context-free grammar by mean of basic analytic combinatorics. Our second result focuses on alignments between two given ordered trees $S$ and $T$. By refining our grammar to align specific trees, we obtain a decomposition scheme for the  space of alignments,  and use it to design an efficient dynamic programming algorithm for sampling alignments under the  Gibbs-Boltz\-mann probability distribution. This generalizes existing tree alignment algorithms, and opens the door for a probabilistic analysis of the space of suboptimal RNA secondary structures alignments.
\end{abstract}

\section{Introduction}

Tree alignments are the natural analog of sequence alignments, and have been introduced by Jiang, Wang and Zhang~\cite{DBLP:journals/tcs/JiangWZ95} to model and quantify the similarity between two (ordered\footnote{In this work, unless explicitly specified, all trees will be rooted and ordered.}) trees. Initially proposed as an alternative to tree-edit distance, the tree alignment model has proven more robust, allowing for the inclusion of complex local operations~\cite{DBLP:journals/tcbb/BlinDDHT10}, and for being generalized to multiple input trees~\cite{Hochsmann:2004:PMR:1024308.1024315}. Consequently, tree alignment has been used in a wide array of applicative contexts, especially RNA Bioinformatics~\cite{Hoechsmann2003}, where RNA secondary structures alignments can be encoded by tree alignments. The minimal cost tree alignment between two trees of size $n_1$ and $n_2$, under classic insertion/deletion/(mis)-match operations, can be computed  using dynamic programming (DP). The current best algorithms have a worst-case time and space complexity respectively in $\mathcal{O}(n_1n_2(n_1+n_2)^2)$ and $\mathcal{O}(n_1n_2(n_1+n_2))$~\cite{DBLP:journals/tcs/JiangWZ95} algorithms, and an average-case time and space complexity (on uniformly drawn instances) in $\mathcal{O}(n_1n_2)$~\cite{DBLP:journals/tcs/HerrbachDD10}. 
 
In the context of sequence alignments, the enumeration of alignments has been the object of much interest in  Computational Biology~\cite{Dress199843,Torres2003,Andrade2014}. Alignments between two sequences over an alphabet $\Sigma$ can be encoded as sequences over an extended alphabet $\Sigma_a$, representing insertions, deletions and (mis)ma\-tches (\textit{e.g.} $\Sigma=\{a,b\}$, $\Sigma_a=\{(a,-),(-,b),(a,b),(a,a),(b,a),(b,b)\}$). Many sequences over $\Sigma_a$ are equivalent if one considers only (mis)ma\-tches of the alignments, \textit{i.e.} they align sequence of same lengths and induce the same sets of matched positions (\textit{e.g.} $(a,-),(-,b)$ and $(-,b),(a,-)$). It is a natural problem to enumerate distinct sequence alignments for two sequences of cumulated length $n$~\cite[pp. 188]{Waterman1995}. Beyond purely theoretical considerations, the decompositions introduced for enumerating distinct sequence alignments were adapted into DP algorithms, \textit{e.g.} for probabilistic alignment based on expectation maximization~\cite{Do2006}, or to compute Gibbs-Boltzmann measures of reliability~\cite{Vingron1990}. 

In the present work, we consider similar questions on \textit{tree alignments}. We are first interested in counting distinct tree alignments, \textit{i.e.} enumerating, up to equivalence, ordered trees whose vertices are labeled in $\Sigma_a$ (called \textit{supertrees} from now). For trees, the notion of equivalence of alignments generalizes that of sequence alignments, \textit{i.e.} two alignments are \textit{equivalent} when they align the same pairs of trees, and induce the same sets of (mis)matched positions. Unfortunately, contrasting with the case of sequence alignments, existing DP algorithms for computing an optimal tree alignment~\cite{DBLP:journals/tcs/JiangWZ95,DBLP:journals/tcbb/BlinDDHT10,DBLP:journals/tcs/SchirmerG13} cannot be easi\-ly adapted into enumeration schemes for tree alignments up to equivalence. This additional difficulty is due to the existence of ambiguities of different nature.

Our main contribution is a grammar for (distinct) tree alignments, which provably generates a single representative for each equivalence class. We use the symbolic method~\cite{flajolet-sedgewick} to obtain the generating function of tree alignments, and asymptotic equivalents for various statistics of interest can easily be derived, such as the average number of alignments over trees of total size $n$. 
Finally, and, perhaps more importantly from an applied point of view, the grammar can be transformed into an unambiguous and complete DP algorithm for aligning two input trees. The resulting algorithm has the same asymptotic worst-case and average-case complexities, up to reasonable constants, as the current best -- ambiguous -- algorithm~\cite{DBLP:journals/tcs/JiangWZ95,DBLP:journals/tcbb/BlinDDHT10}. The main interest of such an algorithm is that it opens immediately the way to new applications for the tree alignment model, including a critical assessment of the reliability of optimal alignments, either obtained by counting co-optimal alignments, or by sampling suboptimal alignments according to a Gibbs-Boltzmann distribution (see~\cite{DBLP:conf/wabi/PontyS11} for an example of this approach for the RNA folding problem). 

In Section~\ref{sec:def} we introduce the main definitions about trees, supertrees and tree alignments. In Section~\ref{sec:grammar}, we provide a grammar that generates all tree alignments. In Section~\ref{sec:enumeration} we analyze this grammar from an enumerative point of view and give precise results on the number of alignments of fixed size. Finally, in Section~\ref{sec:DP} we show how to transform the tree alignments grammar into a dynamic programming algorithm to sample tree alignments between two specified trees.

\section{Definitions}\label{sec:def}

\paragraph{Trees and supertrees.}
Let $\Sigma$ be an \textit{alphabet}. A tree $T$ on $\Sigma$ is a rooted plane tree whose vertices are labeled by elements of $\Sigma$. We denote by $V_T$ the set of vertices of $T$. We \textit{remove a non-root vertex $v$} from a tree $T$ by contracting the edge between $v$ and its parent $u$, that keeps its label. Removing the root $r$ of a tree consists in creating a forest composed of the subtrees rooted at the children of $r$. We denote the operation of removing a vertex $v$ from $T$ by $T-v$.

We denote by $\Sigma_a$ the alphabet defined by $\Sigma_a= \left(\Sigma\cup\{-\}\right)^2 - \{(-,-)\}$. An element $(x,y)\in \Sigma_a$ is an \textit{insertion} (resp. \textit{deletion}, \textit{match}) if $y=-$ (resp. $x=-$, $(x,y)\in \Sigma^2$). A \textit{supertree} $A$ is a tree on $\Sigma_a$; a vertex of $A$ is an insertion (resp. {deletion}, {match}) if its label is an insertion (resp. {deletion}, {match}). The size of a supertree $A$ is the number of its insertions and deletions, plus twice the number of its matches. A \textit{superforest} is an ordered sequence of supertrees.

\begin{figure}[h!]
  \centering
  \includegraphics[scale=1.2]{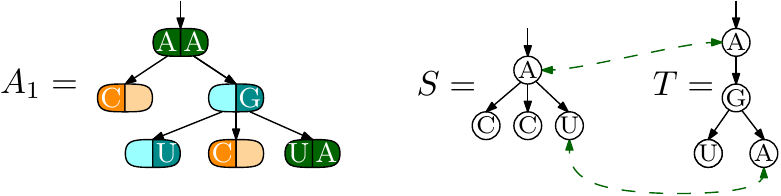}\\
  \caption{A supertree $A_1$ with alphabet $\Sigma = \{A,C,G,U\}$, and the associated trees $S=\pi_1(A_1)$ and $T=\pi_2(A_1)$. 
    The alignment of $S$ and $T$ defined by $A$ is composed of two
    pairs of matched $(A,A)$ and $(U,A)$, indicated by dashed
    arrows. }
  \label{fig:alignment}
\end{figure}

Given a supertree $A$ on $\Sigma$, we define two forests $\pi_1(A)$ and $\pi_2(A)$ as follows: $\pi_1(A)$ (resp. $\pi_2(A)$) is obtained by (1) iteratively removing all insertion (resp. deletions) of $A$, in an arbitrary order, and (2) replacing the label $(x,y)$ of each remaining vertex by $x$ (resp. $y$). We refer to Fig.~\ref{fig:alignment} for an illustration.
We extend the notations $\pi_1$ and $\pi_2$ on vertices: for a non-insertion (resp. non-deletion) vertex $v$ of $A$, we denote by $\pi_1(v)$ (resp. $\pi_2(v)$) the corresponding vertex in $\pi_1(A)$ (resp. $\pi_2(A)$). A vertex $x$ of $\pi_1(A)$ such that $\pi_1^{-1}(x)$ is an insertion (resp. match) is said to be inserted (resp. matched) in $A$. Similarly, a vertex $y$ of $\pi_2(A)$ such that $\pi_2^{-1}(y)$ is a deletion (resp. match) is said to be deleted (resp. matched) in $A$. 

\paragraph{Tree alignments.}
As forests $\pi_1(A)$ and $\pi_2(A)$ are embedded into the supertree
$A$, the latter implicitly defines an \textit{alignment} between the
forests $\pi_1(A)$ and $\pi_2(A)$, \textit{i.e.} a set of
correspondences between vertices of $\pi_1(A)$ and $\pi_2(A)$, that is
consistent with the structure of both
forests~\cite{DBLP:journals/tcs/JiangWZ95}. We refer to Fig.~\ref{fig:alignment} for an illustration.

\begin{figure}[h!]
  \centering
  \includegraphics[scale=1.2]{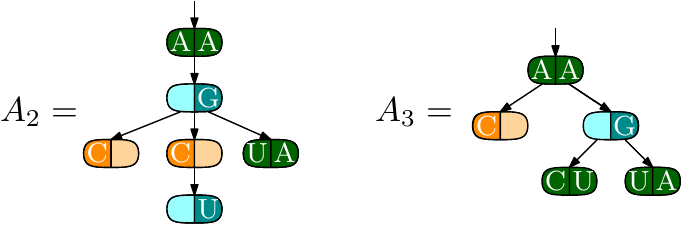}\\
  \caption{Two non-equivalent supertrees, representing two different tree alignments. However, the supertree  $A_1$ from Fig.~\ref{fig:alignment} and the supertree $A_2$ are equivalent.}
  \label{fig:alignment2}
\end{figure}

We now turn to the central notion of \textit{equivalent alignments}, \textit{i.e.} alignments of identical pairs of trees, that contain exactly the same set of matched vertices.  
Given a supertree $A$, representing an alignment between two trees $S=\pi_1(A)$ and $T=\pi_2(A)$, the \textit{set of matches of $A$} is formed by the elements $(x,y)$ of $V_S \times V_T$ such that $\pi_1^{-1}(x)=\pi_2^{-1}(y)$ (\textit{i.e.} there exists a vertex $v$ of $A$ such that $\pi_1(v)=x$ and $\pi_2(v)=y$). Two supertrees $A_1$ and $A_2$ are 
\textit{equivalent} if $\pi_1(A_1)=\pi_1(A_2)$, $\pi_2(A_1)=\pi_2(A_2)$, and the sets of matches of $A_1$ and $A_2$ are identical (see Fig.~\ref{fig:alignment2} for an illustration). 

A \textit{tree alignment} is then defined as an equivalence class over supertrees with respect to the above-defined equivalence relation, for which $\pi_1(A)$ and $\pi_2(A)$ are trees. 
The notion of \textit{forest alignment} is similarly defined when $\pi_1(A)$ and $\pi_2(A)$ are not restricted to trees. 
Given a set $\mathcal S$ of tree (resp. forest) alignments, a set $\mathcal T$ of supertrees (resp. superforests) is said to be \textit{representative of $\mathcal S$} if it contains exactly one supertree (resp. superforest) for each alignment (\textit{i.e.} equivalence classes of supertrees and forests) in $\mathcal S$. 
Tree alignments
will now be the focus of our work.




\section{A grammar for tree alignments}\label{sec:grammar}

In this section, we describe a context-free grammar for a set $\mathcal A$ of supertrees that is representative of the set of all tree alignments. 
\begin{figure}[h!]
\begin{minipage}{\textwidth}
\begin{framed}
  \vspace*{-3mm}
  \begin{eqnarray}
    \mathcal A & \ccsimeq & \Vset \varnothing \oplus \mathcal T_{I} \oplus \mathcal T_{D} \oplus \Insr{\mathcal F_{I} \circ \mathcal T_{D}} \label{eqT} \\
\mathcal T_I & \ccsimeq & \Insr {\mathcal F_I},  \quad \mathcal F_I \ccsimeq \left\{\textrm{empty superforest}\right\} \oplus  \Insr {\mathcal F_I} \circ \mathcal F_I  \\
\mathcal T_D & \ccsimeq & \Insr {\mathcal F_D},  \quad \mathcal F_D \ccsimeq \left\{\textrm{empty superforest}\right\} \oplus  \Insr {\mathcal F_D} \circ \mathcal F_D  \\
    \Vset \varnothing & \ccsimeq  & \Vset \uparrow \oplus \Insr{\VHset}\label{eqVn} \\
    \Vset \uparrow & \ccsimeq  & \Matchr {\Hset \InsDelMode \varnothing \varnothing} \oplus \Delr { \mathcal F_{D} \circ \Vset \uparrow  \circ \mathcal F_{D}} \label{eqVu}
    \\ 
    \VHset  & \ccsimeq & \mathcal F_I \circ {\VHset} \oplus \Vset \varnothing \circ \mathcal F_I  \oplus \Delr { \Hset \InsDelMode \leftrightarrow \varnothing }  \circ \mathcal F_I 
                       \label{eqVH} 
  \end{eqnarray}
 For every $\nu, M, M'$ with $\nu \in \left\{\InsDelMode,\DelMode\right\}$   and  $M,M' \in \{\varnothing,\leftrightarrow,\rightarrow\}$:
  \begin{eqnarray}
    \Hset \nu  {M} {M'} & \ccsimeq & \bigoplus \begin{cases} \left\{\textrm{empty superforest}\right\} &\textrm{if }(M,M') = (\varnothing,\varnothing) \\
      \mathcal T_I \circ \Hset \nu {M} {M'}  & \textrm{if }\nu \neq \ \DelMode \textrm{ and if }M \neq \leftrightarrow \\
      \mathcal T_D \circ \Hset \DelMode { M } {M'} & \textrm{if }M' \neq \  \leftrightarrow \\
      \Vset \varnothing \circ \Hcut M {M'} \one \one &  \\
      \Insr { \Hset \InsDelMode \varnothing \leftrightarrow  }  \circ   \Hcut  M {M'} \one \several    &  \\
      \Delr { \Hset \DelMode \leftrightarrow \varnothing }  \circ  \Hcut  M {M'} \several \one     
    \end{cases}
    \label{eqH}
  \end{eqnarray}
For every $M,M' \in \{\varnothing,\leftrightarrow,\rightarrow\}$ and $i,j \in \{\one,\several\}$:
  \begin{eqnarray}
    \Hcut M {M'} i j & \ccsimeq & \Hset \InsDelMode {\alpha(M)} {\alpha(M')} \oplus \begin{cases} \mathcal F_I &\textrm{ if }M = \varnothing\textrm{ and }M' =\  \rightarrow \\
      \mathcal F_I &\textrm{ if }M = \varnothing,  M' =\  \leftrightarrow\textrm{ and }j = \several \\
      \mathcal F_D &\textrm{ if }M =\  \rightarrow\textrm{ and }M' = \varnothing \\
      \mathcal F_D &\textrm{ if }M =\  \leftrightarrow,\  M' = \varnothing\textrm{ and }i = \several \\
      \varnothing &\textrm{otherwise}
    \end{cases} 
    \label{eqHc}
  \end{eqnarray}
  where $\alpha(\varnothing) = \varnothing$ and $\alpha(\leftrightarrow) = \alpha(\rightarrow) =\ \rightarrow$.
\end{framed}
\end{minipage}
  \vspace*{-0.2cm}
    \caption{A context-free grammar for $\mathcal A$, a representative set of all tree alignments.
  \vspace*{-0.8cm}}
  \label{fig:grammar}

\end{figure}

\begin{figure}[h!]
  \centering
    \includegraphics[width = \textwidth]{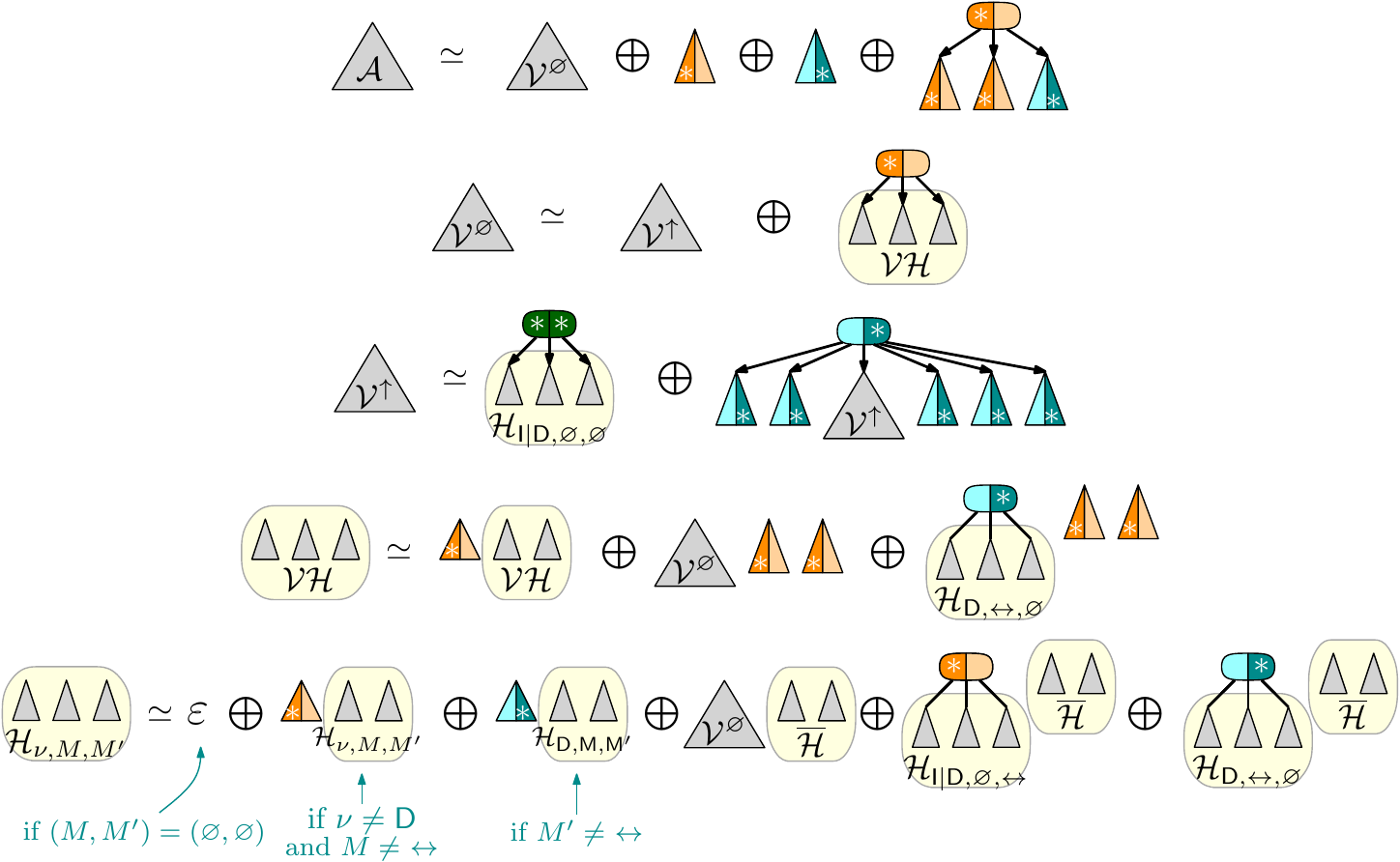}\\
  \vspace*{-3mm}
  \caption{A schematic illustration of the grammar for tree alignments.}
  \label{fig:illustr}
\end{figure}

We first define some basic operations on supertrees and superforests:
\vspace*{-5mm}
\begin{itemize}
\item The (ordered) concatenation of two (super)forests $A$ and $B$ is denoted by $A \circ B$. It creates a new superforest beginning by the supertrees of $A$, and ending by the supertrees of $B$.
\item Given two disjoint sets $\mathcal T_1$ and $\mathcal T_2$ of supertrees or superforests, we denote by $\mathcal T_1 \oplus \mathcal T_2$ their (disjoint) union.
\item For any superforest $A$ and $a,b\in \Sigma$, $\Insrr{A}{a}$ (resp. $\Delrr{A}{b}$, $\Matchrr{A}{a,b}$) denotes the supertree whose root is the vertex $(a,-)$ (resp. $(-,b)$, $(a,b)$) and whose children are the supertrees in $A$, ordered with the same order that they have in $A$. 
\item We naturally extend these operators to a set $\cal T$ of supertrees or superforests:
$\displaystyle\Insr{{\cal T}}=\bigoplus_{A\in {\cal T}, a\in \Sigma} \Insrr{A}{a}$, $\displaystyle\Delr{{\cal T}}=\bigoplus_{A\in {\cal T}, a\in \Sigma} \Delrr{A}{a}$, $\displaystyle\Matchr{{\cal T}}=\bigoplus_{A\in {\cal T}, (a,b)\in \Sigma^2} \Matchrr{A}{a,b}$.
\end{itemize}
Our grammar is described in Fig.~\ref{fig:grammar}, and illustrated in Fig.~\ref{fig:illustr}.

\begin{theorem}\label{thm:grammar}
  The set of supertrees  $\mathcal A$ generated by the grammar \eqref{eqT}-\eqref{eqHc} is representative of the set of all tree alignments; \textit{i.e.} $\mathcal A$ contains exactly one supertree for each equivalence class of supertrees.
\end{theorem}

The key ingredient to prove Theorem~\ref{thm:grammar} stems from the following (semantic) properties for the classes of supertrees and forests that appear in the grammar:
\begin{enumerate}

\item Supertrees in $\mathcal T_I$ (resp. $\mathcal T_D$) contain only insertion (resp. deletion) vertices. 

\item $\mathcal F_I$ (resp. $\mathcal F_D$) is the set of superforests formed by supertrees of $\mathcal T_I$ (resp. $\mathcal T_D$).

\item For $\mu \in \{\varnothing, \uparrow\}$,  $\Vset \mu$ is representative of the set of alignments $A$ with at least one match, such that, if $\mu = \uparrow$, then the root of $\pi_1(A)$ is matched.

\item  $\VHset$ is representative of the set of forest alignments $A$ with at least one match, such that $\pi_2(A)$ is a tree.

\item For $\nu \in \left\{\InsDelMode,\DelMode\right\}$ and  $(M,M') \in \{\varnothing,\leftrightarrow,\rightarrow\}^2$,  $\Hset \nu   M {M'}$ is representative  of the set of superforests $A$ such that
\begin{itemize}
\item if $\pi_1(A)\neq\varnothing$ and $\nu = \DelMode$, then the first tree of $\pi_1(A)$ is matched in $A$;
\item if $M = \rightarrow$,      then the last tree of $\pi_1(A)$ is matched in $A$ (so $\pi_1(A)\neq\varnothing$);
\item if $M' = \rightarrow$,     then the last tree of $\pi_2(A)$ is matched in $A$ (so $\pi_2(A)\neq\varnothing$);
\item if $M = \leftrightarrow$,  then the first and last trees in $\pi_1(A)$ are matched in $A$ (so $\pi_1(A)$ has at least two trees);
\item if $M' = \leftrightarrow$, then the first and last trees in $\pi_2(A)$ are matched in $A$ (so $\pi_2(A)$ has at least two trees).
\end{itemize}

\item For $i,j\in\{\one,\several\}^2$, $\Hcut  M {M'} i j$ is representative of superforests $A'$ such that 
\begin{itemize}
\item there exists a superforest $A$ such that $A \circ A' \in \  \Hset \DelMode M {M'}$;
\item if $i=\one$ (resp. $\several$), $\pi_1(A)$ is a tree (resp. a forest with at least two trees);
\item if $j=\one$ (resp. $\several$), $\pi_2(A)$ is a tree (resp. a forest with at least two trees).
\end{itemize}
\end{enumerate}

These properties can be verified recursively through a tedious analysis of the grammar, and imply quite straightforwardly that $\mathcal A$ contains one and exactly one supertree per equivalence class of supertrees.
\begin{rem}\label{rem:grammar}
For sequences alignments, a grammar generating a representative set of sequence alignments can be easily adapted from the grammar generating all sequences over $\Sigma_a$, \textit{e.g.} by preventing any occurrence to immediately precede an insertion. In the case of trees, the two-dimensional nature of the objects seems to forbid such a simple characterization, and seem to intrinsically mandate intricate combinatorial constructs/grammars. Note however, that our grammar, while complex, remains amenable to efficient computations (Section~\ref{sec:grammarST}). 
\end{rem}
\TODOAll{Should we write a detailled proof sketch in the Appendix?}

\section{Applications}\label{sec:grammarST}
\subsection{Enumerating tree alignments}\label{sec:enumeration}

For the sake of simplicity, we will restrict our attention to $|\Sigma|=1$, \textit{i.e.} the alphabet is restricted to a single letter. The general case follows easily, and will be described in an extended version of the paper.

For a family $\mathcal F$ of superforests, we define a bivariate ordinary generating function 
$$F(t,z)=\sum_{ \substack{ n\geq 0,\, k \geq 0}} f_{n,k} \, t^n \, z^k$$ 
where $f_{n,k}$ is the number of superforests in $\mathcal F$ of size $n$ with $k$ matches.

Using the \textit{symbolic method}~\cite{flajolet-sedgewick}, one classically translates the specification described by Eqs.~\eqref{eqT}-\eqref{eqHc} into a system of functional equations relating the generating functions of the sets of supertrees and forests. To that purpose, classes of objects are replaced by their generating function, disjoint unions (resp. concatenations) of two sets of supertrees are replaced by additions (resp. multiplications) of their generating functions, the addition of a root translates into a multiplication by a monomial $tz$ (resp. $t$) if the root represents a match (resp. insertion/deletion), and empty superforests and sets translate into $1$ and $0$ respectively. The grammar is context-free, so the resulting system is algebraic and can be solved to yield the following characterization result. 

\begin{theorem} The generating functions $T(t,z)$ and $F(t,z)$ of tree and forest alignments,
 whose size and number of matches are marked by $t$ and $z$ respectively, satisfy
\begin{equation}
T(t,z) = \left( t^2 + t - t^2z + \frac {t} {\sqrt{1-4t}}  \right) F(t,z),
\end{equation}
\begin{equation}
(tzC(t)^2 - t^2C(t)^2 + 2t)  F(t,z)^2 + (t^2C(t)^4  - 2t C(t)^2 - 1 ) F(t,z) + C(t)^2 = 0,
\label{quadeq}
\end{equation}
where
$C(t) =   (1-\sqrt{1 - 4t})/2t$ is the generating function of Catalan numbers.
\end{theorem}

Solving the quadratic equation \eqref{quadeq} leads to an explicit formula for $F\!A$ (and hence $T\!A$), details of which are omitted due to space constraints. Nonetheless, these explicit expressions can be used to compute an asymptotic estimate using a \textit{transfer theorem} \cite[Cor. VI.1 p. 392]{flajolet-sedgewick}.

\begin{theorem}\label{thm:enum_alignments}
 The number of tree alignments of size $n$ is asymptotically equivalent to $\kappa\times n^{-3/2}\times 6^n,$ where
$\kappa = \sqrt 2 (3 - \sqrt 3)/(24 \sqrt \pi)$.
\end{theorem}

\begin{cor}\label{cor:enum_alignments} The average number of tree alignments for a random pair of trees of cumulated size $n$ is $\kappa' \times 1.5^n$, where $\kappa' = \sqrt 2 (3 - \sqrt 3)/6$. 
\end{cor}


Similar techniques can be used to characterize the distribution of the number of matches in a random tree alignment. A direct application of \cite[Theorem IX.12 p. 676]{flajolet-sedgewick} indeed gives the following.

\begin{prop}\label{prop:avg_matches} Let $m_n$ be the random variable that counts the number of matches in a uniformly-drawn random tree alignment. The variable $m_n$ follows a Normal law of mean $\mathbb E(m_n) \sim n/6$ and variance $\mathbb V(m_n) \sim n/6$.
\end{prop}



\subsection{Sampling alignments between two given trees}\label{sec:DP}

We now consider two \emph{fixed} trees $S$ and $T$, and consider the task of sampling a tree alignment $A$ such that $\pi_1(A)=S$ and $\pi_2(A)=T$, with respect to the Gibbs-Boltzmann probability distribution. This can be used to assess the stability of a prediction. We refer the interested reader to our introduction for examples of further motivation and possible applications. 

\paragraph{Preliminaries.} Let $\mathcal T_{S,T}$ be the set of all supertrees $A$ such that $\pi_1(A)=S$ and $\pi_2(A)=T$, and $\mathcal A_{S,T}$ be a representative set of $\mathcal T_{S,T}$. In other words, $\mathcal A_{S,T}$ can be interpreted as the set of all alignments between $S$ and $T$. For any supertree $A\in \mathcal T_{S,T}$, we define its \textit{edit score} $s(A)$ as the sum of the number of insertions, deletions and matches $(x,y)$ such that $x\neq y$.\footnote{The present results can be trivially extended to any edit scoring system that is a positive linear combination of the numbers of insertions, deletions and matches.} 

For a given positive constant $k\theta$, the \textit{partition function} $Z_{S,T}$ of $\mathcal A_{S,T}$ and  the \textit{Gibbs-Boltzmann probability} $\PrA(A)$ of an alignment $A\in \mathcal A_{S,T}$ are defined as
$$Z_{S,T}= \sum_{A \in \mathcal A_{S,T}} e^{-s(A)/k\theta}, \quad \PrA(A)=\frac {e^{-s(A)/k\theta}}{Z_{S,T}}.$$
When $k\theta$ tends to $0$, this distribution tends to the uniform distribution over supertrees of minimum edit score, while, when $k\theta$ tends to $+\infty$, it tends toward the uniform distribution over $\mathcal A_{S,T}$.

We consider the following problem: given two trees $S$ and $T$, and a positive constant $k\theta$, design a sampling algorithm for alignments between $S$ and $T$ under the Gibbs-Boltzmann probability distribution. This problem generalizes the classic combinatorial optimization problem of computing a tree alignment between $S$ and $T$ having minimum edit score.

To address this problem, we rely on dynamic programming, by  the approach described, among others, in~\cite{DBLP:conf/wabi/PontyS11} for RNA folding. We begin by adapting the grammar introduced in Section~\ref{sec:grammar} into a grammar for $\mathcal A_{S,T}$,
then detail how this grammar leads to an efficient sampling algorithm.

\paragraph{A grammar for $\mathcal A_{S,T}$.}
In order to guarantee that each supertree $A$ indeed aligns two input trees $S$ and $T$ (namely $\pi_1(A)=S$ and $\pi_2(A)=T$), we need to restrict which rules in the grammar can be used, conditionally to which trees and forests are currently being generated. To that purpose, we introduce, for each set $\mathcal S$ in the previous grammar, an indexed version ${\mathcal S}_{[u,v]}$ which denotes the restriction of $\mathcal S$ to alignments between $u$ and $v$ two forests in $S$ and $T$. 

Slightly abusing previous notations, we denote by $a(u)$ the tree whose root is a vertex $a$ and whose (forest of) children is $u$. Finally, for every tree/forest $X$, $\Ins(X)$ (resp. $\Del(X)$) represents the supertree/superforest obtained from $X$ by inserting (resp. deleting) each of its elements. If $X$ is empty, $\Ins(X)$ and $\Del(X)$ denote the empty superforest.
The grammar for $\mathcal A_{S,T}$ is described in Fig.~\ref{fig:fixed}.

\begin{figure}[h!]
\begin{minipage}{\textwidth}
\begin{framed}
  \vspace*{-3mm}

{\small
\begin{equation}
  \mathcal A_{\substack{S,T\\S\equiv r_S(X_S)}}  \ccsimeq  \Vset \varnothing[S,T] \oplus \Insrr{\mathcal \Ins(X_S) \circ \Del(T)}{r_S}\label{eqT2}
  \end{equation}
  \begin{equation}
  \Vset \varnothing[a(u),b(v)]  \ccsimeq   \Vset \uparrow[a(u),b(v)] \oplus \Insrr{\VHset[u,b(v)]}{a} \label{eqVn2}   
  \end{equation}
  \begin{equation} \hspace*{-3.5mm}
  \Vset \uparrow[a(u),b(v)] \ccsimeq 
  \Union \begin{cases}
    \Matchrr {\Hset \InsDelMode \varnothing \varnothing[u,v]}{a,b} \\
\bigoplus\limits_{Y\circ c(w) \circ Y'=v}\hspace*{-4mm}\Delrr { \Del(Y)\circ \Vset \uparrow[a(u),c(w)]  \circ \Del(Y')}{b} 
  \end{cases}\hspace*{-4mm} \label{eqVu2}
    \end{equation}
  \begin{equation}
    \VHset[\varnothing,b(v)]  \ccsimeq  \varnothing 
      \end{equation}
  \begin{equation}
  \hspace*{-2mm}\VHset[a(u)\circ X,b(v)]   \ccsimeq 
  \bigoplus \begin{cases}
    \Ins(a(u)) \circ {\VHset}[X,b(v)] \\
    \hspace*{-00mm}\bigoplus\limits_{\hspace*{00mm}\substack{X' \circ X''=a(u)\circ X \\ |X'| \geq 2}}
    \hspace*{-4mm}\Delrr { \Hset \InsDelMode \leftrightarrow \varnothing[X',v] }{b}  \circ \Ins(X'') \\
    \Vset \varnothing [a(u),b(v)]  \circ \Ins(X)  
  \end{cases}\hspace*{-5mm}\label{eqVH2}
\end{equation}
\noindent For every $\nu, M, M'$ with $\nu \in \left\{\InsDelMode,\DelMode\right\}$ and  $M,M' \in \{\varnothing,\leftrightarrow,\rightarrow\}$: 
\begin{eqnarray}
\Hset \nu {M} {M'}[X,\varnothing] & \ccsimeq & \begin{cases} \Ins(X) & \textrm{if }(M,M')=(\varnothing,\varnothing), \\ \varnothing & \textrm{otherwise}, \end{cases} \\
\Hset \nu {M} {M'} [\varnothing,Y] & \ccsimeq & \begin{cases} \Del(Y) & \textrm{if }(M,M')=(\varnothing,\varnothing), \\ \varnothing & \textrm{otherwise,} \end{cases}
\end{eqnarray}

\noindent \smallskip$\Hset \nu  {M} {M'}[a(u)\circ X,b(v)\circ Y]  \ccsimeq $
\begin{equation}
 \bigoplus \begin{cases}  \Ins(a(u)) \circ \Hset \nu {M} {M'}[X,b(v)\circ Y]  & \hspace*{-4cm}\textrm{if }\nu \neq \ \DelMode \textrm{ and if }M \neq \leftrightarrow, \\
    \Del(b(v)) \circ \Hset \DelMode { M } {M'}[a(u)\circ X,Y] & \hspace*{-4cm}\textrm{if }M' \neq \  \leftrightarrow, \\
    \Vset  \varnothing [a(u),b(v)] \circ \Hset \InsDelMode {\alpha(M,X)} {\alpha(M',Y)} [X,Y] &  \\
    \bigoplus\limits_{\substack{Y' \circ Y''=b(v)\circ Y \\ |Y'| \geq 2}}
    \hspace*{-5mm} \Insrr { \Hset \InsDelMode \varnothing \leftrightarrow[u,Y']  }{a}  \circ   \Hset \InsDelMode {\alpha(M,X)} {\alpha(M',Y'')} [X,Y'']    &  \\
    \bigoplus\limits_{\substack{X' \circ X''=a(u)\circ X \\ |X'| \geq 2}} 
    \hspace*{-5mm} \Delrr { \Hset \DelMode \leftrightarrow \varnothing[X',v] }{b}  \circ  \Hset \InsDelMode {\alpha(M,X'')} {\alpha(M',Y)} [X'',Y]    
  \end{cases}
  \label{eqH2}
\end{equation}
where  $\alpha(\varnothing,X) = \varnothing$ and $\alpha(\leftrightarrow,X) = \alpha(\rightarrow,X)  = \begin{cases} \varnothing & \textrm{if }X=\varnothing, \\ \rightarrow & \textrm{otherwise.}\end{cases} $
}

\end{framed}
\end{minipage}
  \vspace*{-0.2cm}
    \caption{A grammar for $\mathcal A_{S,T}$, a representative set of all tree alignments between two fixed trees $S$ and $T$.
}
  \label{fig:fixed}
\end{figure}

\begin{theorem}\label{thm:grammarST} 
  Let $S$ and $T$ be non-empty trees.   The set of supertrees $\mathcal A_{S,T}$ generated by grammar \eqref{eqT2}-\eqref{eqH2} is representative of $\mathcal T_{S,T}$ the tree alignments between $S$ and $T$.
\end{theorem}

\paragraph{Applications to dynamic programming.}
The grammar defined by Equations \eqref{eqT2}-\eqref{eqH2} is a decomposition scheme for the alignments between $S$ and $T$. It can easily be transformed into an algorithm for computing the partition function $Z_{S,T}$. Indeed, $Z_{S,T}$ is simply a weighted sum over all possible supertrees of $\mathcal A_{S,T}$, which is a set generated by the grammar. Now consider the image of the grammar as a set of numerical equations, obtained by  syntactically replacing:
\begin{itemize}
\item The operators $(\oplus,\circ)$ with $(\sum,\times)$ respectively;
\item The empty set $\varnothing$ with $0$;
\item Inserted/Deleted trees/forests $\Ins(X)$ and $\Del(X)$ with $e^{-|X|/k\theta}$,
\item Match $\Matchrr{V}{a,a}$ events  with $V$, $\forall a\in \Sigma$ and any expression $V$;
\item Insertion $\Insrr{V}{a}$ events, deletion $\Delrr{V}{a}$ events, and mismatch $\Matchrr{V}{a,b}$ events  with $e^{-1/k\theta} \times V$, $\forall a\neq b \in \Sigma$ and any  $V$.
\end{itemize}
Theorem~\ref{thm:grammarST} immediately implies  that the resulting set is a dynamic programming scheme that computes $Z_{S,T}$ instead of $\mathcal A_{S,T}$. 

Moreover, each non-terminal term of the modified grammar now contains the partition function of the set of supertrees associated to this non-terminal term in the set-theoretic grammar, \textit{\textit{e.g.}} a term $\VHset[a(u)\circ X,b(v)]$.
This information can then be used to define an algorithm to sample supertrees from $\mathcal A_{S,T}$ under the Gibbs-Boltzmann distribution, following the \textit{recursive} method for random generation~\cite{Wilf1977}.  

To do so, it suffices to reinterpret the grammar defined by Equations \eqref{eqT2}-\eqref{eqH2} as a branching process: each $\oplus$ operator is replaced by a branching operator that, instead of joining sets of supertrees into a larger set of supertrees, chooses one of the sets according to the weight of its partition function. For instance, assume we have a grammar rule $U \ccsimeq V \oplus W$:  the sampling algorithm will select one of the sets $V,W$, with $V$ being chosen with probability $Z_V/(Z_V+Z_W)$, and $W$ with probability $Z_W/(Z_V+Z_W)$, provided that $Z_V,\ Z_W$ and $Z_X$ have been previously computed. Recursive calls will then result into a supertree, which is provably randomly generated under the Gibbs-Boltzmann distribution.

\begin{theorem}\label{thm:samplingST}
Let $S$ and $T$ be two trees of respective sizes $n_S$ and $n_T$. The above-defined branching process adapted from grammar \eqref{eqT2}-\eqref{eqH2} defines an algorithm that samples a supertree from $\mathcal A_{S,T}$ under the Gibbs-Boltzmann distribution. The worst-case time and space complexities of the algorithm are in $\mathcal O(n_S\,n_T\,(n_S+n_T)^2)$, while the average-case time and space complexities are in $\mathcal O(n_S\,n_T)$.
\end{theorem}

The correctness of the algorithm immediately follows  from Theorem~\ref{thm:grammarST}. Its complexities are identical to ~\cite{DBLP:journals/tcs/JiangWZ95,DBLP:journals/tcs/HerrbachDD10} since the structure of the DP scheme essentially remains the same; only the number of DP tables is increased (by a constant factor). This implies that our algorithm, while solving a much more general problem, retains the same asymptotic complexity (up to constants) than the current tree alignment algorithms that are limited to computing a single optimal tree alignment.

\TODOAll{Average-time complexity? Or do-we keep it for the full version} 

\section{Conclusion and discussion}\label{sec:conclusion}
 
Following a classical line of research in string algorithms, we introduced the notion of equivalence for tree alignments, and described a context-free grammar for a representative set of all possible alignments. We also showed how this grammar can be used to derive asymptotic properties of alignments, and design an efficient dynamic programming sampling algorithm for alignments between two given trees. 

From an applied point of view, our results allow to sample optimal, as well as suboptimal, tree alignments for a pair of given trees under the Gibbs-Boltzmann distribution; following the program outlined in~\cite{DBLP:conf/wabi/PontyS11}, we are currently using this algorithm to revisit the alignment of RNA structures.

Our proposed grammar for tree alignments is more complex than the grammars used to generate a representative set of sequence alignments, although dynamic programming for computing optimal sequences and trees alignments are very similar. This is due to the fact that it is particularly hard to characterize a representative set of tree alignments (see Remark~\ref{rem:grammar}). It thus remains an open problem to design a representative set of tree alignment that would be amenable to enumeration using a simpler grammar. However, it is important to remark that, despite its apparent complexity, our grammar leads to algorithms with an asymptotic complexity of the same order than existing optimization algorithms. 

From a theoretical point of view, we believe that tree alignments as defined in this work form an interesting combinatorial family whose properties deserve to be explored in depth. More generally, it would be interesting to characterize the conditions under which an instance-agnostic grammar, enumerating a search space, could be adapted into a decomposition for a specific instance. Such a theory, at the confluence of enumerative combinatorics and algorithmic design, could provide another principled ways to design dynamic-programming algorithms.

\TODOCedric{
Summary (maybe when article is a bit more mature?)\\
Outline differences with sequence alignment, both in term of nature of gen. fun. and in average-case behavior for total input length $n$. Average-case complexity in $n^2$, similar to previous ambiguous propositions and, while intriguing, very probably an artifact of asymptotic average properties of trees (running the algorithm on two forests takes $\Theta(n^3)$)\\
Future applications of a DP scheme, falls within the hypergraph framework, and should allow for extensions such as inside/outside, Max expected accurary alignment...
}

\bibliographystyle{splncs03}
\bibliography{tree_alignment,DP}
  
\ShowTODO{
\newpage
\appendix
\setcounter{tocdepth}{5}

\listoftodos} 
\end{document}